\documentclass[10pt,times,a4paper]{elsarticle}
\usepackage[utf8]{inputenc}
\usepackage{amsmath,graphicx}
\usepackage{amsfonts}
\usepackage{amssymb}
\journal{Giornale di Fisica}
\let\today\relax
\makeatletter
\def\ps@pprintTitle{%
    \let\@oddhead\@empty
    \let\@evenhead\@empty
    \def\@oddfoot{\footnotesize\itshape
         {S. Lepri et. al Giornale di Fisica, VOL. LXIV, N. 3 p.275 (2023) (english translation)} \hfill\today}%
    \let\@evenfoot\@oddfoot
    }
\makeatother

\begin{document}
\begin{frontmatter}

\title{1953: Fermi's ``little discovery" \\ and the  
birth of the numerical experiment}
\author[lepri1]{Stefano Lepri}
\author[livi,lepri1]{Roberto Livi}
\author[ruffo,lepri1]{Stefano Ruffo}

\address[lepri1]{Consiglio Nazionale delle Ricerche, Istituto dei Sistemi Complessi, Via Madonna del Piano 10 I-50019 Sesto Fiorentino, Italy} 
\address[livi]{Dipartimento di Fisica e Astronomia, 
Universit\`a di Firenze, Via G. Sansone 1, I-50019 \\ Sesto Fiorentino, Italy} 
\address[ruffo]{SISSA Via Bonomea 265, I-34136 Trieste, Italy}

\begin{abstract}
The year 1953 is pivotal for computational physics:
the first application of the Monte-Carlo method is published and  
calculations of the so-called Fermi-Pasta-Ulam-Tsingou 
experiment are started.
It is the beginning of the massive use in the physical sciences of numerical methods 
implemented on electronic computers and a decisive step in the development of modern nonlinear
dynamics. This will lead to an 
unpredictable development during the following 70 years. 
We briefly review the unfolding of these events and present some 
recent results that show how the issues raised are still relevant today.
\end{abstract}

\end{frontmatter}

\section{A step back in time}

In 1953, the Los Alamos National Laboratory was a complex of buildings in the New Mexico desert. The facility had been established ten years earlier as a \textit{top secret} research center  for the development of nuclear weapons during World War II, and was still heavily involved in that task.
Let's put ourselves in the shoes of a hypothetical visitor, describing his or her 
experience:

\begin{quotation}

As I walked through the dusty streets of the lab's main campus, I saw scientists and engineers hurrying back and forth, carrying stacks of papers and computer printouts. Many of them wore thick glasses and lab coats, and all of them seemed focused and serious, as if they had the weight of the world on their shoulders.

Everywhere I looked, there were signs of the lab's cutting-edge research. In one building, I glimpsed a group of scientists peering through microscopes, studying the structure of atoms. In another, I saw a group of engineers huddled around a massive machine, making adjustments to its gears and wires.

But perhaps the most impressive sight was the lab's newest creation: the Mathematical Analyzer, Numerical Integrator, and Computer, or MANIAC for short. This massive machine, which took up an entire room, was one of the fastest and most powerful computers in the world. Its thousands of vacuum tubes and punch-card programming made it an essential tool for the nuclear weapons lab's work.

Despite the seriousness of the lab's work, there was also a palpable feeling of excitement and possibility in the air. Los Alamos was at the forefront of a new era of science and technology, and the scientists and engineers working there knew that their work had the potential to change the world in profound ways.

As I left the lab and returned to the desert, I could not help but feel a sense of wonder at the incredible work being done behind those walls. Los Alamos was a place of innovation, determination and possibility, and I knew that its work would continue to shape the world for years to come.
\end{quotation}

Very suggestive. But even more suggestive 
to know that the ``testimony" was generated by asking the
now-famous 
ChatGPT to narratively describe Los Alamos
in 1953. Apart from some naivete (such as studying the structure
of atoms by peering into a microscope of the time) what better way to emphasize the upheaval 
that the advent of computers and software has brought in
(only) 70 years? The world has really changed as profoundly as
presumably the scientists envisioned by artificial intelligence expected.

Even limiting ourselves to (our) specific field of Theoretical Physics
the progress is truly impressive: as every researcher knows 
well, the possibility of simulating a model of given
physical system is a very powerful
tool. And we are not only referring to the possibility
to deal with an enormous amount of data, but above all to devise
\textit{Gedankenexperiments} (ideal experiments) aimed at understanding a 
phenomenon and perhaps, as each of us hopes in our
daily work, to discover a new one.

Well, all this originated precisely in the dusty streets
of Los Alamos 70 years ago when, for the first time, 
people thought of using the impressive computational tools
as real conceptual laboratories. 
It is therefore with good reason that one considers what took place in those
years as the birth moment of computational physics or of the
\textit{experimental mathematics} \cite{dauxois2005fermi,porter2009fermi}.
And it is in the year 1953 that two crucial steps take place:
the publication of the first 
article where the use of the Monte-Carlo technique is introduced and demonstrated
and the first calculations of the famous
numerical experiment conceived by Fermi, Pasta and Ulam.

On the occasion of this anniversary, we aim here to provide 
a brief historical account of the experiment by Fermi and collaborators 
and an equally brief update of the ideas that continue to flow
from it. The relevance of the problem and approach 
has been emphasized in so many publications, including those 
directed to a wider audience 
\cite{falcioni2001contributo,porter2009fermi,de2016modello}.
We will therefore be concise, referring to the literature 
for more specific issues.  We begin by introducing the protagonist of these
scientific enterprises.

\begin{figure}
\includegraphics[width=0.8\textwidth]{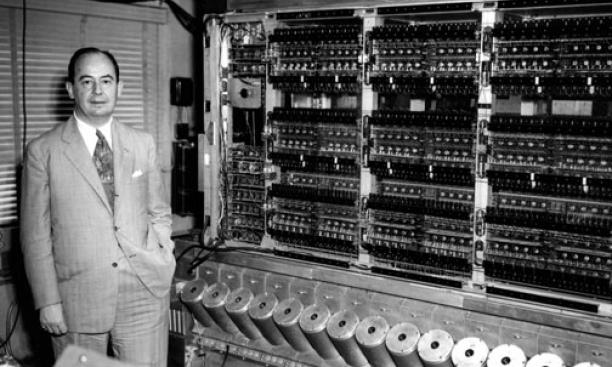}
\caption{John von Neumann in front of MANIAC in 1952. The calculator
was over 20 tons heavy and took up an entire room. It consisted of thousands of vacuum tubes, which were used to perform logical operations, and was programmed using punched cards.
}
\label{fig:1}       			  
\end{figure}

\section{MANIAC}

John von Neumann,  inspired by the work of Alan Turing,  is considered to be 
the mind behind the machines of the MANIAC family (fig. \ref{fig:1}).
According to Metropolis \cite{metropolis1990alamos} 
they were both convinced that
\begin{quotation}
modern computers could have profound effects in nonlinear mathematics by virtue of observing the unfolding of mathematical solutions that could stimulate thoughts about models and structures in those very refractory areas.
\end{quotation}

\begin{figure}
\includegraphics[width=0.7\textwidth]{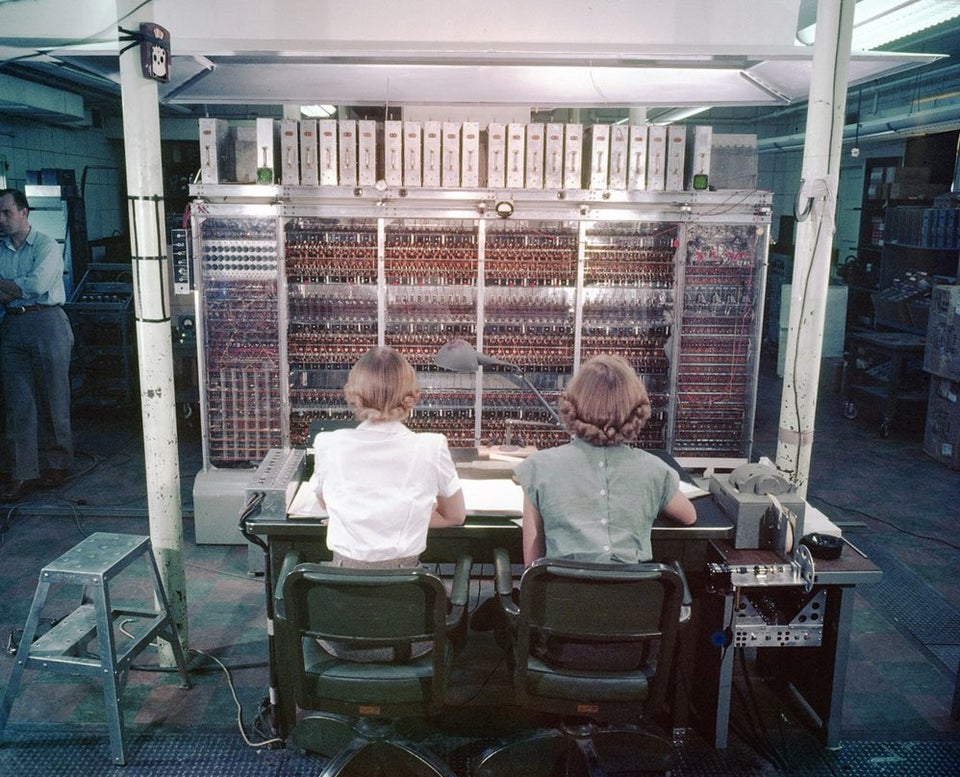}
\caption{MANIAC I in 1952 in color!
}
\label{fig:2}       			  
\end{figure}

The internal architecture of today's computers is, in fact, that conceived
by von Neumann.  His intention was that MANIAC would initiate a computer revolution, enabling the solution of scientific problems that had been deemed
impossible. To maximize the impact of such a program, he avoided applying for patents and published detailed reports on its progress -- a spirit that today we would call of 
\textit{open science}.  The development 
of the first machine at the Institute for Advanced Study in Princeton
had already begun in 1946.
Soon, seventeen computers were built 
according to its specifications, including the MANIAC in Los Angeles, identically named
to that of Los Alamos, and IBM's first machine.
Since it was a copy of the one in Princeton, the implementation 
of the Los Alamos MANIAC I proceeded apace and, after a 
development phase, the first \textit{run} 
took place in March 1952 (fig. \ref{fig:2}) \cite{metropolis1990alamos}.

Despite its size and complexity, MANIAC was surprisingly fast, capable of performing thousands of calculations per second.  Obviously, 
by today's standards this is laughable performance. You have to think
that the entire computer memory (five kilobytes) is equivalent to an amount of memory less than 
that occupied by a single
icon on our laptops or cell phones.  Also
no one had yet invented a modern programming language:
the development of the first FORTRAN compiler, the 
\textit{number crunching} language \textit{par excellence}, 
began shortly thereafter, in 1954 \cite{maynard2012daybreak}.

Among its many uses, the machine seemed particularly suitable for solving 
of differential equations that described novel problems, often
(unfortunately) motivated by wartime research.
As Nicholas Metropolis, one of the leading figures of that  
pioneering season, recalls
\cite{metropolis1990alamos} 
\begin{quotation}
But the war was to change all that by demanding solutions to all kinds of nonlinear differential equations. At Los Alamos, from the very beginning in April 1943, it was clear that nonlinearities would come into their own --- a new experience for the theorists who had learned to touch only linear phenomena and had a built-in shyness for anything beyond.
\end{quotation}
In this respect, the MANIAC manual states that
\footnote{\texttt{www.bitsavers.org/pdf/lanl/LA-1725\_The\_MANIAC\_Jul54.pdf}} 
\begin{quotation}
The differential equations are of such complexity that analytical
methods are not known for obtaining their solutions. The only recourse
is to numerical procedures; therefore these problems are ideally suited
for the computer.
The first step in the solution of the problem is to replace the 
differential equations by a set of finite-difference equations. We do not
discuss here the stability or convergence of such methods, but only mention
them as necessary considerations in writing the difference equations.
\end{quotation}

So the new possibilities were truly remarkable, and the theorists soon overcame 
their \textit{shyness} toward nonlinear problems.  One example is the
Monte-Carlo method which, although already known to statisticians, 
started with the possibility of generating sets of pseudo-random numbers.
And indeed, 1953 is also the year of publication of the famous article
"Equation of State Calculations by Very Fast Computing Machines" di N.
Metropolis, A. W. Rosenbluth e M. N. Rosenbluth, e M. Teller e E. Teller 
\cite{metropolis1953equation}.  
Its abstract states that 
 \textit{Results for the two‐dimensional rigid‐sphere system have been obtained on the Los Alamos MANIAC and are presented here.} 
 Such work
marked the beginning of the use of the Monte-Carlo method for problem solving
in the physical sciences. The method described in this publication later became known as
the Metropolis algorithm, undoubtedly the most famous and most widely used Monte Carlo algorithm, as evidenced by the impressive number of citations that 
approaches 50000 to date.  Interestingly, none of the authors 
of the paper made subsequent use of the algorithm, 
and they remained unknown to the simulation physics community that has developed since this publication.  
Their role in the development of the algorithm became a subject of mystery and legend, 
as recounted in 
\cite{gubernatis2005marshall}.
The impact of Monte-Carlo methods on the physical sciences was 
undoubtedly enormous, finding applications in all fields, from  
particle physics to condensed matter.

Later, in 1956, MANIAC I had a further moment of glory 
to the general public as well, 
becoming the first computer to beat a human being in a 
game of chess.  It was actually a 
simplified version of the game (a 6-by-6 board, without bishops) 
known as \textit{Los Alamos Chess}.  To that 
result  contributed also Stanislaw (Stan) Ulam \cite{kister1957experiments}, 
one of the protagonists of the story 
we will now discuss.

\section{The Los Alamos Report LA-1940}          
\label{sec:foo}

The results of the numerical experiment are reported in a now famous 
report \cite{Fermi1955}, which has never been published in a scientific journal. 
The work was conceived in the summer of 1952 and
the calculations were carried out in the following summer.
The model studied, which from the name of the authors, Enrico Fermi, John Pasta and Stan Ulam 
(fig. 3) has long been referred to by the acronym FPU,  
is presented as the discretization of a one-dimensional continuum
with the extremes fixed containing nonlinear terms. When the number of equations 
tends to infinity, the system tends to the well-known wave equation
\footnote{In this respect, it is curious to recall that in 1918 Fermi passed
brilliantly the entrance examination to the Scuola Normale in Pisa precisely
with a dissertation entitled
\textit{Caratteri distintivi dei suoni}
where he writes \textit{si trova l'equazione differenziale della 
verga vibrante, risolta per mezzo dello sviluppo in serie 
di Fourier, dopo aver trovato le autofunzioni e gli autovalori del
problema} [the differential equation for a vibrating bar is found 
and solved by Fourier series, after having determined the eigenfunctions 
and eigenvalues of the problem] \cite{segre2020enrico}. }  
with nonlinear terms  \textit{of a complicated nature}. 
In the original notation, the equations thereby considered are,  
denoting by $x_n$ the displacement 
\begin{equation}
\ddot x_n= x_{n+1}-2x_n+x_{n-1}+\alpha[( x_{n+1}-x_n)^2-( x_{n}-x_{n-1})^2]
\end{equation}
where $n=1,\ldots, N$ ($N=64$ in the report) and  
\begin{equation}
\ddot x_n= x_{n+1}-2x_n+x_{n-1}+\beta[( x_{n+1}-x_n)^3-( x_{n}-x_{n-1})^3],
\end{equation}
that are respectively called “model $\alpha$" and “model $\beta$". 
Fixed boundary are enforced letting $x_0=x_{N+1}=0$.
The total mechanical energy is conserved and the associated Hamiltonian
has the general form
\begin{equation}
H = \sum_{n=1}^N \frac{p_n^2}{2} +  \sum_{n=0}^N U(x_{n+1} - x_n)\,,
\label{Hamil}
\end{equation} 
where 
\begin{equation}
U(x) \equiv \frac{1}{2} x^2+\frac{\alpha}{3} x^3 + \frac{\beta}{4} x^4.
\label{eq:fput}
\end{equation}
This formulation includes both model $\alpha$ ($\alpha\neq 0,\beta=0$) and $\beta$
($\alpha=0, \beta\neq0$),  but also the so-called
“$\alpha\beta$ model" ($\alpha\neq 0,\beta\neq 0$), 
where both nonlinear terms are present.

Considering the case of fixed boundary conditions, the \textit{normal modes } of the problem are defined by the transformations that 
diagonalize the quadratic part of the Hamiltonian,
\begin{equation}
Q_k=\sqrt{\frac{2}{N+1}}\sum_i x_i \sin\left(\frac{\pi k i}{N+1}\right)\quad
P_k=\sqrt{\frac{2}{N+1}}\sum_i \dot x_i \sin\left(\frac{\pi k i}{N+1}\right),
\end{equation}
and the main observables are the normal modes energies 
($k=1,\ldots,N$)
\begin{equation}
E_k=\frac{1}{2}P_k^2 + \frac{\omega_k^2}{2}Q_k^2, \quad 
\omega_k\equiv 2\sin\left(\frac{\pi k }{2(N+1)}\right).
\end{equation}

The numerical experiment consists in choosing
initial conditions strongly out of equilibrium, in which all the energy is assigned to one or 
few normal modes of large wavelength, and in monitoring 
the energy distribution of the modes over time.
Fermi's idea was to test on what time scale 
energy was transferred from the macroscopic to the microscopic scale, 
reaching equilibrium.  More precisely, the state of equilibrium corresponds 
to the equipartition of energy among the normal modes, that is, to the situation in which
each mode, on average, shares the same fraction
of the total energy, initially attributed to the initial condition. 
The question then is whether the time-averaged $E_k$ converges
to the equilibrium value prescribed by the energy equipartition principle
and, if so, on what time scale.

\begin{figure}[h]
\includegraphics[height=0.26\textwidth,clip]{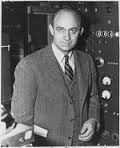}
\includegraphics[height=0.26\textwidth,clip]{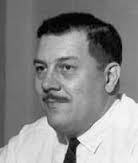}
\includegraphics[height=0.26\textwidth,clip]{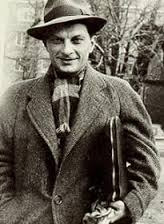}
\includegraphics[height=0.26\textwidth,clip]{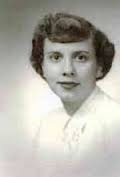}
\caption{Enrico Fermi, Jim Pasta, Stan Ulam and Mary Tsingou Mentzel.
}
\label{fig:3}    
\end{figure}
\subsection{“...features which were, from the beginning, surprising to us"}

To the Authors'  surprise, on the time scale accessible to MANIAC,
the equilibrium state \textit{was not} reached. 
The result is stated in the
summary, which concludes by summarizing the results concisely and effectively:
\begin{quotation}
The results show very little, if any, tendency toward equipartition of energy 
among the degrees of freedom.
\end{quotation}

\begin{figure}
\includegraphics[width=0.8\textwidth]{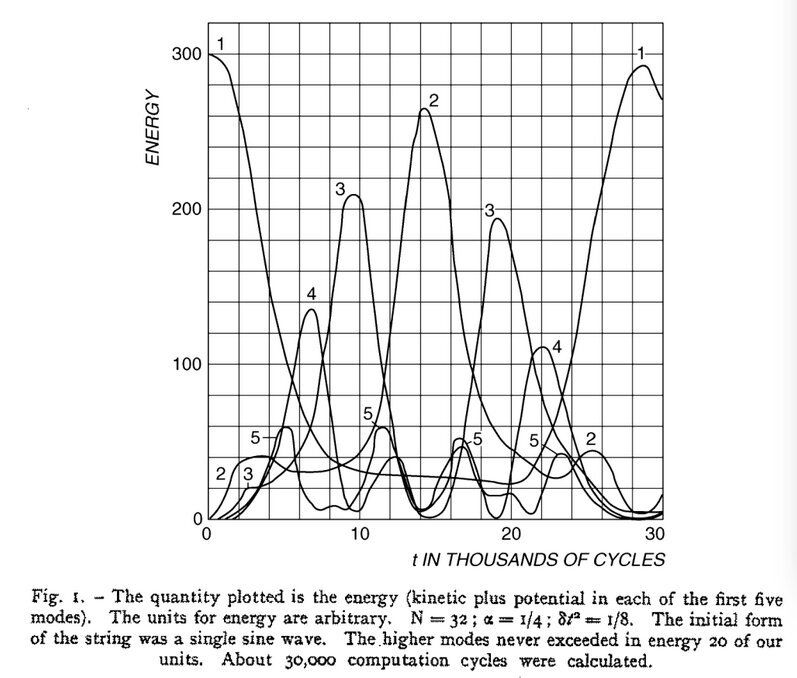}
\caption{Figure 1 of the original report showing the  
time evolution  of the energies of the first 5 normal modes starting
from a single sine-wave initial condition.
}
\label{fig:4}       			  
\end{figure}

Rather, a (quasi)steady state is observed, in which only some normal modes 
are excited.  In addition, the phenomenon of \textit{recurrence} is observed.
that is, the energy of a mode returns to a value very close to its initial value, 
as visible in the famous figure reproduced in fig.\ref{fig:4}.
This apparent paradox has been the source of many studies
and we will discuss some recent developments in a moment. But first we must
talk about another protagonist in the story.

\subsection{“We thank miss Mary Tsingou..."}

Another excerpt from the  MANIAC user manual:
\begin{quotation}
There are several phases to the solution of
a problem by an electronic computer.
First, there is the formulation of the problem itself
by the mathematician or theoretical physicist.
Second, this is followed
by the detailed preparation of the problem by the programmer for the
specific computer.
\end{quotation}

It was therefore necessary for someone to translate the problem
into a flowchart and implement the design into a code 
usable by the machine.  Fermi himself had shown keen interest
in programming already back in the 40s when, with his
characteristic approach, he self-taught  the necessary techniques 
\cite{metropolis1990alamos}.  Indeed, 
as testified by Ulam himself \cite{Fermi1955}, 
Fermi learned very
quickly to program MANIAC and worked personally on the
flowcharts and the preparation 
of the code. 
A footnote to the report thanks
Mary Tsingou \textit{for efficient
coding of the problems and for running the computations on 
the Los Alamos MANIAC machine.}
In the era when scientific journals demand to specify 
each author's contributions to an article and it is not considered
rightly permissible to exclude contributors,  
this seems unconceivable.  Even more so considering
that we are talking about a young woman.  
\footnote{As a matter of fact,  
several 
female researchers were employed by the staff of MANIAC I, 
as listed by Table I in \cite{metropolis1990alamos}.}

 In recent times, Mary Tsingou's role has been 
widely recognized \cite{dauxois2008fermi,dauxois2020dis}, 
and it has become common to replace the acronym FPU with
FPUT. Although belated, a due recognition to those who 
have spent a great deal of time and energy to achieve the
results, laboriously turning flowcharts of the 
code into punch cards, with all the difficulties
that anyone who had to deal with the procedures 
of \textit{debugging} can figure out.  

In this regard, female researchers played a major role.   On the occasion of this anniversary it is appropriate to mention 
among them Arianna W. Rosenbluth, considered  to be the author of the implementation
of the Metropolis algorithm and the first person to implement the 
Monte-Carlo method
\cite{metropolis1953equation}.
  
\subsection{"After the untimely death of Professor E. Fermi .."}

Let us come back to the report: at the end of the summary we learn that after the 
Fermi's passing in 1954, computational activity continued.
What would Enrico Fermi, the \textit{Papa} as he was nicknamed 
in Via Panisperna, have thought, from the height of his boundless knowledge of
physics, of the developments of his \textit{litte discovery}?
The profound influence of the "never published article" \cite{falcioni2001contributo} cannot be underestimated. Just to give a rough idea of its significance let
us mention the following items:

\begin{itemize}
\item As mentioned at the beginning, it demonstrates an innovative use of the computer
as a tool for investigation through what today we would call
\textit{numerical experiments}.
\item It may be considered one of the starting points of the 
\textit{molecular dynamics}, the approach that allows the calculation of
dynamical and statistical  properties
based on numerical integration 
of the microscopic equations of motion. This approach 
has numerous applications in physics, and is
complementary to Monte-Carlo methods, which originated in precisely the 
same context.  \footnote{For a detailed and comprehensive historical approach to the
birth and impact of this important discipline on theoretical physics, see 
the recent volume \cite{battimelli2020computer}}.
\item Has motivated the study of \textit{deterministic chaos} and its role
in securing one of the foundations of statistical mechanics 
namely the ergodic hypothesis, bringing this problem back to the attention of physicists.
\item The attempt to explain the recurrence paradox led
to the modern theory of \textit{nonlinear wave equations} (Boussinesq, KdV...) and
of integrability for nonlinear systems with many degrees of freedom.
\end{itemize}

We refer the reader to  \cite{ford1992fermi,carati2002fermi,gallavotti2007fermi,campbell2005introduction}
for an account of the subsequent developments. 
In what follows 
we summarize some more recent ones, without claiming to be 
complete,  with the aim of showing that this is a still an active field of
research .
\section{The shadow of Toda}
\label{sec:toda}

From the very beginning, thanks to the contribution of Zabusky and Kruskal, it had been assumed that 
the phenomenon of recurrences originated from the quasi-integrability
of the model.  More recently, a new point of view has emerged
\cite{benettin2013fermi}.  The new viewpoint is actually based on an 
old knowledge, namely the Hamiltonian of the form (\ref{Hamil}) with the 
exponential potential
\begin{equation}
U_{Toda} = U_0\left(e^{\lambda x} -1 -\lambda x\right).
\label{toda}
\end{equation}
Morikazu Toda formulated it in 1967 
(and has since then
beared his name) and determined its analytical solutions.   Subsequently 
\cite{henon1974integrals,flaschka1974toda} it was proven to 
have a special property:
it is one of the rare examples of a \textit{nonlinear integrable} system, which 
have as many constants of motion as there are degrees of freedom. As is known 
from analytical mechanics, by means of a canonical transformation to
action-angle variables $I,\varphi$, the corresponding Hamiltonian can be rewritten
as a function of actions only, $H(I)$. Thus, the dynamics turns out to be 
quasi-periodic, remaining confined on a multi-dimensional torus.

\begin{figure}
\includegraphics[width=0.7\textwidth]{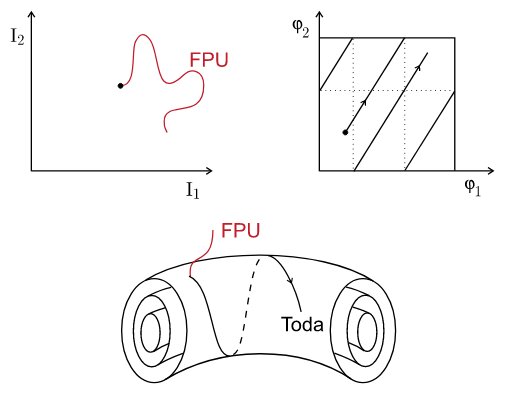}
\caption{The motion of an integrable system, such as the Toda model,  
occurs on an invariant torus in which the actions are 
constant and the angles grow linearly with time. 
The presence of terms that break integrability
induces a slow motion in the transverse direction,
but mantain the trajectories close to the torus. 
On this time scale the dynamics is practically 
indistinguishable from that of the underlying integrable system:
This results in  a metastable, 
non-ergodic state (Courtesy of G. Benettin
and A. Ponno).}
\label{fig:toro}       			  
\end{figure}

The main point can be summarized simply:
Toda's model represents an integrable approximation of FPUT 
more accurate than the harmonic chain. 
Thus, the slow chaotic motion leading to equipartition
is not so much due to the fact that FPUT is a discretization of
 an integrable wave equation, but rather to the fact that 
the dynamics is
 (at least at low enough energies) essentially indistinguishable
 from Toda's on very long time scales 
 (fig. \ref{fig:toro}).  This viewpoint has been established only 
 recently and represents a significant paradigm shift 
 in explaining the problem, even in quantitative terms
 through scaling laws \cite{benettin2020understanding}.
Furthermore, the metastable state can be seen as 
 a generalized Gibbsian state, the generalization 
 to the integrable case of the 
 canonical state of usual statistical mechanics
\cite{goldfriend2019equilibration}.

\section{Further developments}
 
The amount of research problems generated 
by the FPUT numerical experiment is so vast that it cannot be
completely accunted for in this brief article.  In this section
we report a selection of some of these issues, referring readers
to the cited literature.

\subsection{The Anti-FPUT  problem}

The original numerical experiment considered initial conditions 
focused on the normal modes of large wavelength. What happens if instead
we concentrate the initial energy on those at the other end of the spectrum,
namely the modes of wavelengths comparable to the lattice spacing?
In this case, through a mechanism known as  \textit{modulational instability},
localized excitations are spontaneously formed that 
correspond to so-called  \textit{discrete breathers} or, in the limit 
of small amplitudes, \textit{envelope solitons} \cite{kosevich2000modulational,dauxois2005anti}. 
The thermalization process is determined by the interaction
of this "gas" of breathers that collide and exchange 
energy in a peculiar way: the excitation tends to concentrate over 
time in a single breather that is eventually destroyed by the 
fluctuations. The system relaxes to equipartition,
on time scales that increase as energy density decreases.


 \subsection{The mathematics of FPUT}

In 1954, a year before the Los Alamos report, Kolmogorov announced in Amsterdam one of the most important results in twentieth-century mathematics. That theorem, now known as KAM (from the names of the three mathematicians credited with its authorship: Kolmogorov himself, Arnold and Moser), provides sufficient conditions for the "integrable" behavior of a Hamiltonian system to survive for all times even when the system is perturbed. However, the applicability of the KAM Theorem to the FPUT model has only recently been demonstrated. The first positive result, valid only for specific chain lengths and only for the $\beta$ model, was due to Nishida in 1971. It took another 30 years before the applicability of this result was proved first for the $\beta$  model \cite{rink2006proof} and then also for the other models \cite{henrici2008results} . As a consequence of applying the KAM theorem, we now know that, when the number of particles in the chain is fixed, in a low-energy regime the system will never reach thermodynamic equilibrium. 

Following the route  opened by Zabusky and Kruskal regarding the link between the Korteweg de Vries (KdV) equation and FPUT, the first mathematically rigorous results date back to the early 2000s. In \cite{bambusi2006metastability} it was shown that in an appropriate continuous limit and after a "normal form" step, the FPUT model is described by a pair of KdV equations describing waves traveling in opposite directions.
As for the connection with Toda's model, the first to realize it was Manakov in 1974, followed by a very important work by Flashka, Ferguson and McLaughlin \cite{ferguson1982nonlinear}. 
But to arrive to two theorems linking the length of equipartition times in FPUT and Toda dynamics we have to wait for the work \cite{bambusi2016birkhoff}. Here, the authors show that if the energy of the system is low enough, the Toda actions are very close to the normal modes of the chain.  Thus the quasi-conservation of Toda actions implies lack of equilibration for the FPUT model on the timescale for which the approximation holds, in agreement with the numerical results. Only very recently in 
\cite{maiocchi2014averaging} and in \cite{grava2020adiabatic} the authors manage to show that, if the temperature is low enough, Toda actions are adiabatic invariants for the FPUT system even in the thermodynamic limit.

Finally, we mention another approach to the FPUT problem based on the 
wave-kinetic (or wave turbulence) method that has been recently 
proposed, see \cite{onorato2023wave} for an account.

\subsection{Almost-integrability and transport}

The quasi-periodic recurrence of initial states 
discovered by FPUT not only challenged spontaneous evolution toward thermodynamic equilibrium, but was also incompatible with the presence of any diffusive mechanism for energy transport.

One might therefore wonder what happens if the FPUT model 
is subjected to an external thermodynamic force, such as a 
temperature gradient. In a normal transport regime, we would expect it to
to behave like an ordinary thermal conductor for
which the well-known Fourier's law states that the heat flow 
passing through it is proportional to the temperature gradient itself.
Instead, it turns out that this relationship is not verified and that 
the system behaves as a "heat superconductor", in the 
sense that the thermal conductivity diverges in the thermodynamic limit \cite{lepri1997heat}.  This phenomenon of so-called \textit{anomalous transport}, although not 
directly related to the slow thermalization described above, represents another 
surprise of the model and has been studied in detail, even experimentally
\cite{benenti2023non}. 

But the problem becomes even more intriguing in the limit of 
low temperatures where FPUT dynamics is, as described, well approximated 
by the Toda chain.  In the latter, solitons propagate freely 
and give rise to ballistic transport, characteristic of  
independent quasi-particle.  From the point of view described 
in section \ref{sec:toda}, that is, considering the FPUT model as
a weak perturbation
of Toda, near the integrable limit, the quasi-particles interact 
and acquire a finite but large mean free path $\ell$. 
Therefore, if the chain size $L$ is such that $L<\ell$ purely ballistic behavior is still observed, whereas if $L>\ell$ the solitons 
diffuse giving rise to normal transport.
Thus Toda's shadow manifests again, restoring 
Fourier's law at least in a regime of intermediate sizes.  
But this does not last forever: at even higher $L$ sizes, one recovers
the anomalous behavior mentioned above \cite{lepri2020too}, 
and the effects of quasi-integrability are lost in the chaos of 
hydrodynamic fluctuations.

\section{D{\`e}j{\`a} vu in the lab}

There is also some experimental evidence of the physics of FPUT,
particularly of the phenomenon of recurrence 
\cite{van2001experimental,wu2007experimental} which, suggestively, 
has been defined as the analogue of the \textit{D{\'e}j{\`a} vu} in the
inanimate world \cite{akhmediev2001deja}. Nonlinear optical systems
are good examples.  For example, the propagation of the
electric field in an optical fiber is accurately described 
by the nonlinear Schr\"odinger equation.
\begin{equation}
i \psi_z +\psi_{xx}+2\psi|\psi|^2=0
\label{nls}
\end{equation}
ruling the evolution of the envelope $\psi(x,t)$ 
of a quasi-monochromatic field 
in a medium with weak optical nonlinearity ($z$ represents the direction
of propagation).
The equation is fully integrable and also has solitonic solutions.
In addition, the plane wave-like solutions are modulationally unstable, and if the 
perturbation consists, for example, of a single mode, after growing to a certain maximum amplitude, it reduces in size and eventually disappears.  From this point of view, 
the recurrence of FPUT can be seen as is a periodic solution of the wave equation (\ref{nls}) \cite{akhmediev2001deja}. 
On the basis of these considerations, several 
experimental observations of the recurrences in various optical systems
have been demonstrated
 \cite{pierangeli2018observation,goossens2019experimental,vanderhaegen2020observation}.
 
\section{Conclusion}

A first conclusion we would like to emphasize is how models, 
however mathematically simple, can lead to new 
and surprising discoveries and, as in the case of FPUT, even 
open up new fields of research, even interdisciplinary ones. 

Moreover, before concluding, it seems to us that this is the appropriate place to emphasize
the contribution of the Italian physics community to the 
FPUT. Even from the brief bibliography provided here, it is clear that many of the 
works on the topic, since its inception, are signed by Italian researchers 
(Rome, Milan, Padua, Florence) who continue 
to be passionate about the problem and provide new advances,
both theoretical and experimental.

Finally, as we have seen, the many current developments in the problem originate
from the creative use of the computer as a tool for investigation.
We began this article with the help of the most modern
and powerful algorithmic resource of today. Indeed, we are witnessing a new 
and impressive 
technological breakthrough that will surely impact on the methods
of scientific research, as well as on the rest of the society:
\begin{quotation}
[...] there is now a new and very real possibility for a novel synthesis of advances in experimental technique, high-performance computing, and theory: automatically building theories from data. That is,to the extent we understand pattern, we can use machines to find emergent organization in our vast data sets. And from there, the machines can build our theories, most likely with guidance from a new generation of theorists. \cite{crutchfield2014dreams}
\end{quotation}
Understanding the potential and methodological implications 
offered by such an evolution, in a creative and original approach as 
happened at Los Alamos 70 years ago, could pave the way for new and unexpected
scientific advances.

\bibliography{bibfpu}
\bibliographystyle{acm}
\end{document}